\documentclass[aps,twocolumn, superscriptaddress, svgnames]{revtex4-2}
\usepackage[T1]{fontenc} 
\usepackage{tikz,tikzscale, relsize, pgfplots, xcolor, amssymb, amsmath}
\usetikzlibrary{backgrounds, arrows, calc, decorations, positioning, decorations.pathmorphing,decorations.pathreplacing, 
	decorations.markings, fixedpointarithmetic, fit, patterns, decorations.text, backgrounds,matrix,plotmarks, spy}
\usepgfplotslibrary{fillbetween}
\pgfplotsset{compat=1.17}
\usetikzlibrary{external}
\usepackage{graphicx}
\usepackage{subcaption}
\usepackage{ragged2e}

\begin{document}
\title{Transport of Intensity Phase Retrieval in the Presence of Intensity Variations and Unknown Boundary Conditions}

\author{A. Lubk}
\email{a.lubk@ifw-dresden.de}
\affiliation{Leibniz Institute for Solid State and Materials Research Dresden, Helmholtzstraße 20, 01069 Dresden, Germany}
\affiliation{Institute of Solid State and Materials Physics, TU Dresden, Haeckelstraße 3, 01069 Dresden, Germany}
\author{R. Kyrychenko}
\email{r.kyrychenko@ifw-dresden.de}
\affiliation{Leibniz Institute for Solid State and Materials Research Dresden, Helmholtzstraße 20, 01069 Dresden, Germany}
\author{D. Wolf}
\affiliation{Leibniz Institute for Solid State and Materials Research Dresden, Helmholtzstraße 20, 01069 Dresden, Germany}
\author{M. Wegner}
\affiliation{Leibniz Institute for Solid State and Materials Research Dresden, Helmholtzstraße 20, 01069 Dresden, Germany}
\author{M. Herzog}
\affiliation{Leibniz Institute for Solid State and Materials Research Dresden, Helmholtzstraße 20, 01069 Dresden, Germany}
\author{M. Winter}
\affiliation{Max Planck Institute for
Chemical Physics of Solids, 01187 Dresden, Germany.}
\affiliation{Dresden Center for Nanoanalysis, cfaed, Technical University Dresden, 01069 Dresden, Germany.}
\affiliation{Leibniz Institute for Solid State and Materials Research Dresden, Helmholtzstraße 20, 01069 Dresden, Germany}
\author{O. Zaiets}
\affiliation{Leibniz Institute for Solid State and Materials Research Dresden, Helmholtzstraße 20, 01069 Dresden, Germany}
\affiliation{Leibniz Institute for Solid State and Materials Research Dresden, Helmholtzstraße 20, 01069 Dresden, Germany}
\author{P. Vir}
\affiliation{Max Planck Institute for
Chemical Physics of Solids, 01187 Dresden, Germany.}
\author{J. Schultz}
\affiliation{Leibniz Institute for Solid State and Materials Research Dresden, Helmholtzstraße 20, 01069 Dresden, Germany}
\author{C. Felser}
\affiliation{Max Planck Institute for
Chemical Physics of Solids, 01187 Dresden, Germany.}
\author{B. B\"uchner}
\affiliation{Leibniz Institute for Solid State and Materials Research Dresden, Helmholtzstraße 20, 01069 Dresden, Germany}
\affiliation{Institute of Solid State and Materials Physics, TU Dresden, Haeckelstraße 3, 01069 Dresden, Germany}


\begin{abstract}
The so-called Transport of Intensity Equation (TIE) phase retrieval technique is widely applied in light, x-ray and electron optics to reconstruct, e.g., refractive indices, electric and magnetic fields in solids. Here, we present a largely improved TIE reconstruction algorithm, which properly considers intensity variations as well as unknown boundary conditions in a finite difference implementation of the Transport of Intensity partial differential equation. That largely removes reconstruction artifacts encountered in state-of-the-art Poisson solvers of the TIE, and hence significantly increases the applicability of the technique.
\end{abstract}

\maketitle
\section{Introduction}

A large class of sensing techniques employing optical or matter (e.g., electron) waves requires the retrieval of the wave's phase to obtain crucial information about the matter or fields that have interacted with the wave. For instance, when a fast electron beam employed in transmission electron microscopy (TEM) transmits a specimen, the phase of the electron wave is modulated by static magnetic and electric potentials via elastic scattering \cite{Lubk2018c}. Therefore, in various types of microscopy, such as optical microscopy or electron microscopy, methods have been developed that retrieve the phase (or phase differences) from the intensity distributions recorded on a detector \cite{Zernike1942,Gabor(1948)a}. In the following, we focus on the so-called transport of intensity equation (TIE) phase retrieval method \cite{Teague(1983)}. It is one of the most widely applied schemes because it requires comparably little experimental effort and works for arbitrary sample geometries (see, e.g., Ref. \cite{Zuo2020} for a comprehensive introduction). The TIE method reconstructs the phase from two or three mutually slightly defocused microscopy images by solving an elliptic partial differential equation --\, the TIE. In the context of TEM --\, the application case considered in this work --\, the TIE method has been predominantly used to reconstruct magnetic fields in nanomagnets, e.g., pertaining to skyrmions \cite{Yu2010,Schneider2018, Jena2022}, but also electrostatic fields \cite{Beleggia(2004),Ishizuka(June2005)} (see again Ref. \cite{Zuo2020} for a comprehensive overview containing also numerous examples from, e.g., visible light, X-ray, neutron and atom optics, e.g., for biological and medical imaging). 

Notwithstanding, the TIE method in its predominantly employed form suffers from a number of fundamental challenges that often introduce reconstruction artifacts and hence degrade the information obtained from the phases: (I) Intensity variations of the wave functions are frequently either neglected or considered in an incomplete way by equating the phase with the scalar "potential" pertaining to irrotational (particle) currents \cite{Teague(1983),Paganin(1998),Ishizuka(June2005)}. This derives from the argument that a plane wave elastically interacting with a pure phase object (i.e., a specimen that only modulates the phase) does not exhibit intensity variations if the object plane is stigmatically imaged on the detector. In practical applications, such as the 3D reconstruction of magnetic fields by vector field electron tomography \cite{Phatak2010,Yu2022}, however, amplitude variations due to dynamical scattering and scattering absorption at bend and inhomogenoeus samples of varying thickness are ubiquitous. (II) The boundary conditions (BCs) required to solve the TIE are often not known and strongly deviate from homogeneous Dirichlet, Neumann, or periodic BCs. Notwithstanding, the most frequent scheme assumes periodic BCs for numerical reasons \cite{Gureyev1997}, but also homogeneous Dirichlet or Neumann BCs are employed for their simple implementation \cite{Volkov(2002),Gureyev(2003)}. The latter corresponds to a vanishing current across the boundary, which may hold in situations, where the sample is surrounded by a field-free environment within the reconstruction domain. Depending on the phase structure of the wave, the impact of using the wrong BCs may degrade to some degree with the distance from the boundary (see examples below). (III) The reconstruction of small spatial frequencies in the phase is mildly (i.e., algebraically) ill-conditioned and, hence, requires regularization to suppress error amplification \cite{Paganin(1998)}. This typically amounts to applying a high-pass filter rendering TIE inapt for the reconstruction of slowly varying phase features \cite{Mitome2009}. Indeed, the above three challenges are intertwined in practical applications, because, e.g., erroneous long-range phase variations due to erroneous BCs or neglected amplitude variations are suppressed by regularization at the expense of introducing a regularization error. We note furthermore that partial coherence poses a challenge of its own, as it violates the framework of the TIE. Its implications will not be treated here.

The above challenges have been only partially addressed in previous studies, and no general solution addressing all of them and leading to a practically usable algorithm has emerged so far. For instance, a finite element solution of the TIE \cite{Lubk(2013)a, Parvizi2015,Zhang2020} or dedicated iterative algorithms\cite{Zuo2014a}, properly taking into account intensity variations and allowing arbitrary but fixed predefined BCs did not incorporate a determination of the true BCs. Sophisticated regularization schemes beyond high-pass filtering have been developed for constant intensity cases only.\cite{Bostan2016} 

In this letter, we demonstrate a finite difference scheme that (A) takes into account intensity variations, (B) allows retrieving the correct BCs through a variational scheme, and (C) can be additionally regularized, if necessary. Indeed, the need for regularization is drastically reduced / removed by (A) and (B) in the considered examples pertaining to the reconstruction of magnetic fields in micron- and nanoscale magnetic textures. 

\section{Finite Difference Solution to TIE}

The stationary paraxial wave equation 
\begin{equation}
\frac{\partial\Psi\left(\mathbf{r},z\right)}{\partial z}=\frac{i}{2k_0}\Delta\Psi\left(\mathbf{r},z\right) \label{eq:paraxial_wave}
\end{equation}
is valid for a large variety of scattering phenomena of, e.\,g.\ electrons and photons,  moving within a small solid angle around the optical axis ($z$-axis) \cite{Schmalz(2011), Lubk2018c}.
Here, $k_0$ is the wave number, $\Delta$ the 2D Laplace operator, and $\mathbf{r}=\left(x,y\right)^\top$ the 2D position vector. The TIE scheme is based on the equation of continuity of the paraxial wave equation \cite{Teague(1983)} 
\begin{align}
\frac{\partial\rho\left(\mathbf{r},z\right)}{\partial z} &=-\frac{1}{k_0}\nabla\cdot\boldsymbol{j}\left(\mathbf{r},z\right) \label{eq:TIE}\\
&=-\frac{1}{k_0}\nabla\cdot\left(\rho\left(\mathbf{r},z\right)\nabla\varphi\left(\mathbf{r},z\right)\right) \nonumber
\end{align}
with $\boldsymbol{j}$ being the 2D (particle) current, $\nabla=\left(\partial_{x},\partial_{y}\right)^\top$ the 2D gradient, and $\rho=\left|\Psi\right|^{2}$ the particle density (= image intensity). In order to recover the phase $\varphi$ one records an in-focus image and two defocused images\,--\,$\rho\left(z-\delta z\right)$ and $\rho\left(z+\delta z\right)$, where the wave optical defocus corresponds to propagation along $z$\,--\,and approximates
\begin{equation}
\frac{\partial\rho\left(z\right)}{\partial z}=\frac{\rho\left(z+\delta z\right)-\rho\left(z-\delta z\right)}{2\delta z}+\mathcal{O}\left(\delta z^{2}\right)\label{eq:drhodz}\,.
\end{equation}
Indeed, experimental noise and artifacts (e.g., small image shifts) impose a lower limit to the defocus value $\delta z$ in order to detect a significant intensity change, which, on the other hand, increases the $\mathcal{O}\left(\delta z^{2}\right)$ error, i.e., deviations from the infinitesimal TIE intensity variations. Further below, we exploit the finite defocus to obtain information about the unknown BCs.

 In a simply connected domain (i.e., $\rho>0$ everywhere) the TIE  may be then solved by providing suitable BCs on the boundary of the reconstruction domain \cite{Gilbarg(1983)}. It simplifies to the Poisson equation if $\rho=$const. A Poisson equation is also obtained when seeking the irrotational part of the (particle) current $\boldsymbol{j}=\nabla \phi$, which may be used to characterize (partially) incoherent beams \cite{Paganin(1998),Lubk(2015)}. The phase associated with this irrotational current, however, deviates from the true phase of the wave function in a non-local way if rotational currents are present (see Appendix \ref{app: IrrCur}) \cite{Schmalz(2011), Zuo2014, Lubk(2013)a}. The Poisson equation may be solved very efficiently in Fourier space assuming periodic BCs, which is used in most reconstruction algorithms \cite{Gureyev1997, Allen(2001)} (see also Appendix \ref{app: FDTIE}). Homogeneous pseudo Dirichlet and Neumann BCs can be incorporated into this scheme by duplicating the reconstruction domain appropriately \cite{Volkov(2002)}. A Poisson equation solver with homogeneous pseudo Dirichlet BCs is used below as a reference (see Appendix \ref{app: IrrCur} for further details on the various Poisson-type approximations of the TIE).

In the following, we numerically solve the TIE using a finite difference scheme: We interlace the equidistant $x$ and $y$ sampling points of the 2D image intensity so that the phase and intensity at the sampling points may be represented as vectors. Subsequently, we approximate the directional derivatives along $x$ and $y$ by the left- and right-sided difference to the nearest neighbours, which can be written as a multiplication with bidiagonal matrices. With these building blocks, the whole TIE may be written as a system of linear equations, which can be inverted to yield $\varphi$ (see Appendix \ref{app: FDTIE} for details). In order to suppress error amplification of the (mildly) ill-conditioned TIE at large range phase variations (small spatial frequencies), the linear equation system may be additionally regularized by a Tikhonov scheme as illustrated in the Appendix \ref{app: FDTIE}. In order to incorporate arbitrary inhomogeneous Dirichlet BCs, we embed the sampled discrete phase image in a frame comprising one pixel. The values on this frame are then assigned to Dirichlet BCs (see again Appendix \ref{app: FDTIE} for details). We finally note that the above system of equations may be solved in a numerically efficient way only if exploiting the sparse character of the involved matrices. Thus, all matrices are represented by sparse representations in the code and sparse equation solvers are employed (see \cite{TIE} for the algorithm).

\section{Variation of Boundary Conditions}

The paraxial wave equation (\ref{eq:paraxial_wave}) may be mapped to the coupled system of the TIE and the quantum Hamilton Jacobi (QHJ) equation by representing the wave function by amplitude and phase. Consequently, the TIE does not contain the whole paraxial wave dynamics. Indeed, the focal series inline holographic reconstruction scheme may be linked to integrating the QHJ equation when using small focal steps, hence being complementary to TIE in a certain sense \cite{Lubk(2015)a}. Consequently, the finite propagation distance (defocus) $\delta z$ introduces intensity modulations, which may be described only by propagating the full wave function and not the TIE. We exploit this for determining the unknown BCs by setting up the following variational scheme. We first define the cost function:
\begin{align}
    L &= \int_D|\rho\left(\boldsymbol{r},z+\delta z \right)-\|\Psi_\mathrm{TIE}\left(\boldsymbol{r},z+\delta z\right)\|^2|\\
    &+|\rho\left(\boldsymbol{r},z-\delta z \right)-\|\Psi_\mathrm{TIE}\left(\boldsymbol{r},z-\delta z\right)\|^2|dxdy\,, \nonumber
\end{align}
which computes the difference between the experimental defocused image intensities and the reconstructed defocused intensities. Here, the TIE reconstructed wave function at defocused planes
\begin{equation}
 \Psi_\mathrm{TIE}\left(\boldsymbol{r},z\pm\delta z\right)|=\hat{T}_\mathrm{F}\left(\boldsymbol{r},\pm\delta z\right)*\sqrt{\rho\left(\boldsymbol{r},z\right)}e^{i\varphi_\mathrm{TIE}\left(\boldsymbol{r},z\right)}   
\end{equation}
is obtained by convoluting ($*$) the in focus reconstructed wave with the Fresnel propagator $\hat{T}_\mathrm{F}$ (see Appendix \ref{app: FresnelProp} for details). This cost function is minimized by optimizing the BCs employing non-linear optimizer algorithms. Here, we restrict the numerical complexity (and hence improve convergence) by optimizing only a few low Fourier coefficients (i.e. we assume smooth slowly varying BCs). Moreover, we have to employ global optimizers (which may be augmented by local ones) such as DIRECT-L\cite{Gablonsky2001} because the above cost function can contain several local minima.

\section{Behavior of Finite Difference TIE Algorithm}

In this section, we examine the behavior of the algorithm using simulated phase modulation of a fast electron wave after transmitting two  magnetic structure examples -- a Landau pattern in a square disk with a magnetic vortex and antivortex at its center -- that are frequently occurring in magnetic thin film geometries, easy to model (see Appendix \ref{app: LandauPattern} for details) and suited to demonstrate the main features of the enhanced TIE reconstruction, namely consideration of intensity variations and automatic determination of BCs.

\begin{figure}
\centering
\includegraphics[width=1\columnwidth]{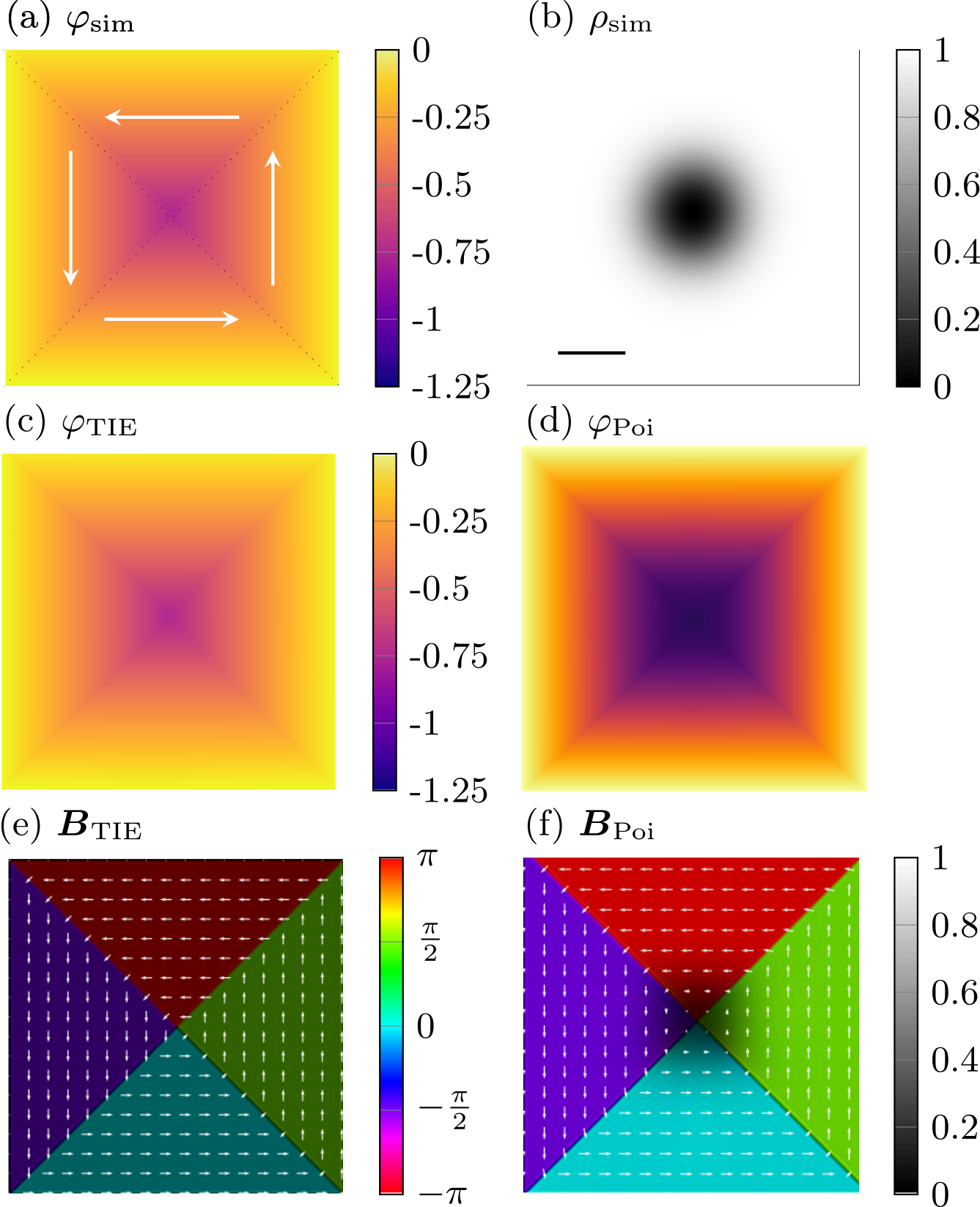}
\caption{\justifying TIE reconstruction of magnetic vortex Landau pattern. (a) Phase and (b) bright-field intensity of simulated electron wave. (c,d) Reconstructed phase using TIE (c) and Poisson algorithm (d).  Reconstructed projected magnetic fields from (e) TIE and (f) Poisson algorithm. The field angle is color-coded, the field magnitude normalized to maximum of $\boldsymbol{B}_\mathrm{Poi}$ is depicted as brightness. Superimposed arrows show the field at a sparser grid. The scale bar in (b) is 0.2\,$\mu$m}
\label{fig:vortex}
\end{figure}

In the predefined vortex, the magnetic Aharonov-Bohm phase picked up the electron wave after transmitting the sample increases linearly from the center toward the boundary in the 4 homogeneous domains (see Appendix \ref{app: LandauPattern}). The domain boundaries are assumed to be sharp (within the modelled spatial resolution) and hence the phase is constant at the boundaries and is deliberately set to zero (Fig. \ref{fig:vortex}a). This allows us to abstract from the issue of general inhomogenoeus BCs and to address the problem of intensity variations. To include intensity variations, we imprinted a radial Gaussian intensity damping on the beam (Fig. \ref{fig:vortex}b). Subsequently, we calculated the slightly defocused intensity before applying TIE phase reconstruction by numerical propagation.

The results of the reconstruction evidence that the phase obtained from the finite difference TIE solution (Fig. \ref{fig:vortex}b) reproduces the original phase, whereas the phase resulting from the Poisson method (Fig. \ref{fig:vortex}c) deviates from the original within the region of non-zero intensity variation in that intensity variations erroneously spill over in the reconstructed phase. Consequently, the magnetic induction obtained from directional derivatives of the reconstructed TIE phase also reproduce the original (Fig. \ref{fig:vortex}e), whereas magnetic inductions from the Poisson method come out too large and exhibit an artificial damping in the region of the intensity damping (Fig. \ref{fig:vortex}f).

\begin{figure}
\centering
\includegraphics[width=1\columnwidth]{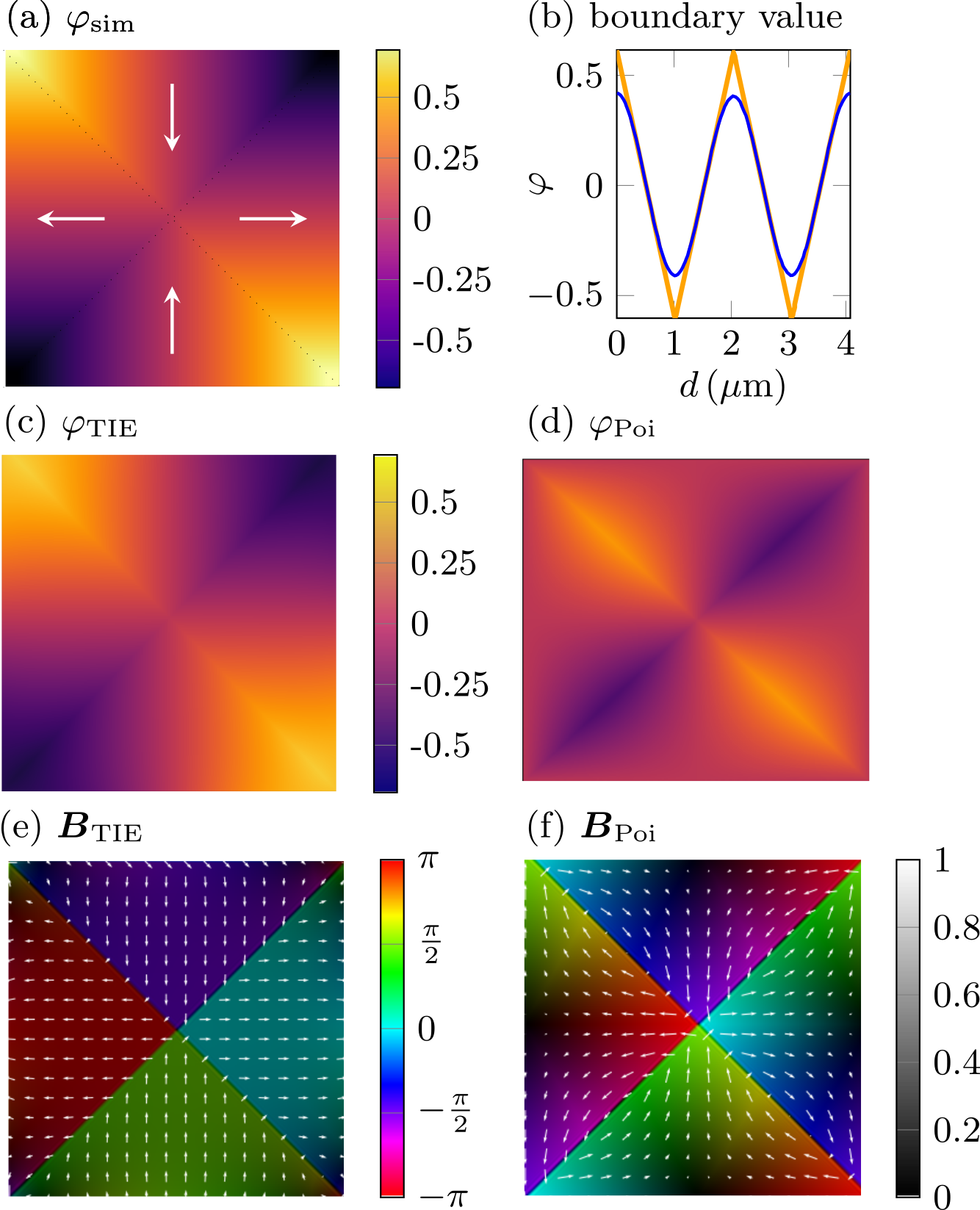}
\caption{\justifying TIE reconstruction of antivortex Landau pattern. (a) phase of simulated electron wave. Domains and domain walls are indicated. (b) boundary value of simulated (blue) and reconstructed (orange) phases. Reconstructed phase from (c) TIE and (d) Poisson algorithm.  Reconstructed projected in-plane components of the magnetic field from (c) TIE and (d) Poisson algorithm. The magnetic field angle is color-coded, the field magnitude normalized to maximum of $\boldsymbol{B}_\mathrm{Poi}$ is depicted as brightness. Superimposed arrows show the field at a sparser grid.}
\label{fig:antivortex}
\end{figure}

In the next case study, the predefined antivortex differs from the vortex in the domain arrangement (Fig. \ref{fig:antivortex}a), as can be established under certain circumstances in a softmagnetic thin film  \cite{Kloodt2018}. Consequently, the magnetic phase at the boundary is not constant and increases and decreases in a zigzag fashion, which corresponds to non-homogeneous Dirichlet BCs. No intensity variations of the beam have been incorporated in this theoretic example in order to focus on the effect of non-trivial BCs. The phase reconstructed from the finite-difference TIE (Fig. \ref{fig:antivortex}b) again closely reproduces the original phase, where deviations only occur at positions, where the domain walls cross the reconstruction domain boundary (corners of the reconstruction domain). These deviations can be traced back to the limited number of Fourier coefficients (4 in this case) of the boundary function recovered by the variational optimization, which is not suited to reproduce a zigzag function at the cusps (see Fig. \ref{fig:antivortex}(b)). The Poisson algorithm on the other hand fails to reconstruct the antivortex phase at the boundary, while the phase structure at the center closely resembles the origin (Fig. \ref{fig:antivortex}(c)). Therefore, the magnetic fields reconstructed from the finite difference TIE also reproduce the original domain pattern, whereas the Poisson reconstruction exhibits strong erroneous magnetic induction rotations and magnitude variations.

\section{Experimental Results}

\begin{figure}
\centering
\includegraphics[width=1\columnwidth]{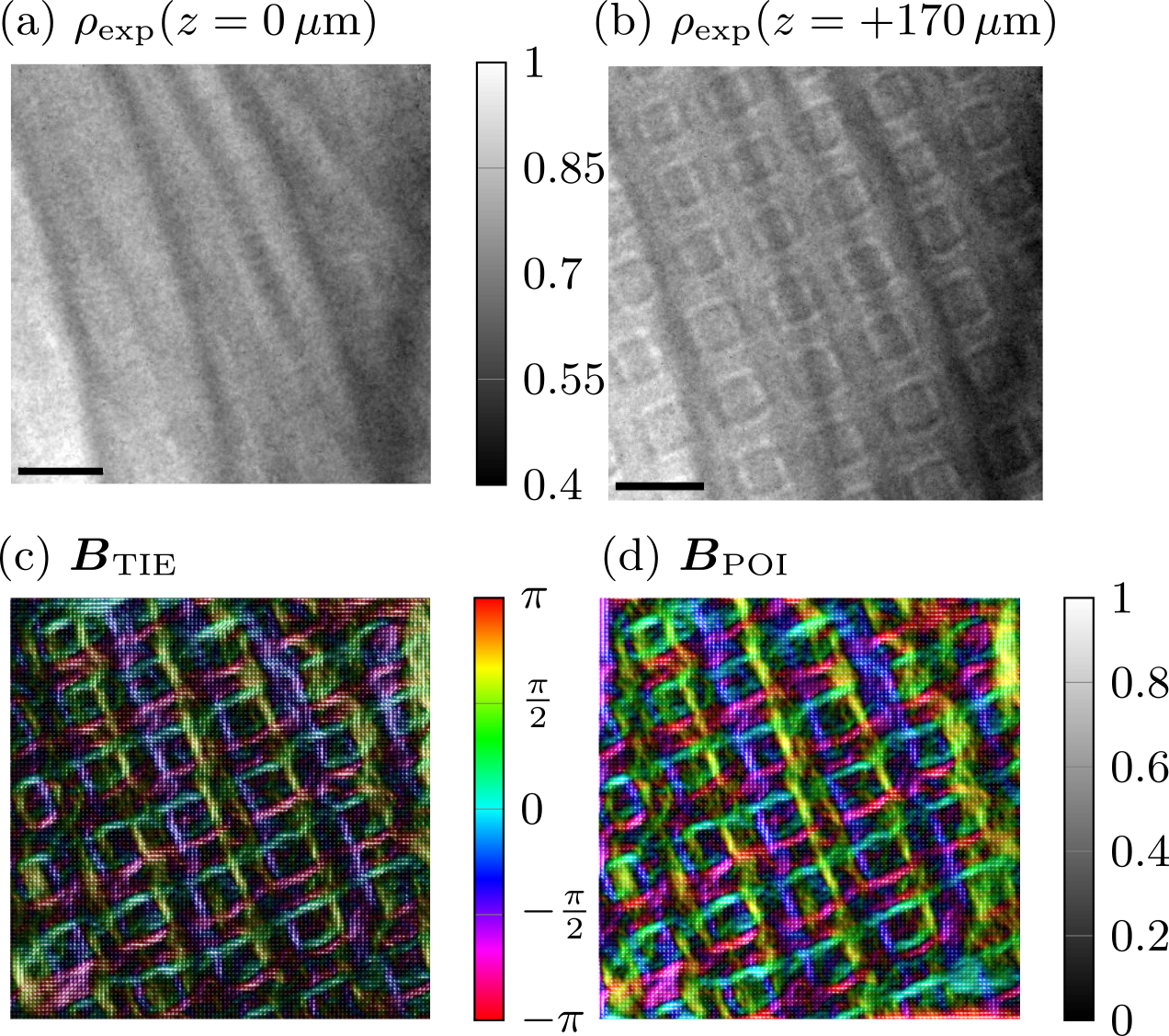}
\caption{\justifying TIE reconstruction of antiskyrmion lattice in Mn$_{1.4}$PtSn. LTEM images at (a) in focus and (b) overfocus of 170 $\mu$m; (c) magnetic field reconstructed with TIE; (d) magnetic field reconstructed with Poisson method. Both reconstructions are regularized with $\lambda=125.4$ corresponding to a reciprocal high pass of radius 20\,${\mu}\mathrm{m}^{-1}$. Scale bars correspond to 0.2 ${\mu}$m. The gray colormap encodes the image intensity (a,b) and the normalized magnetic field magnitude (c,d), the hsv colormap encodes the angle of the magnetic field vectors (c,d).}
\label{fig:MnPtSn}
\end{figure}

We finally test the improved TIE algorithm on two experimental examples, which are subject to experimental noise and span a variety of typical challenges. We will compare the improved TIE reconstruction with the Poisson solution and abstain from an in-depth analysis of the reconstructed magnetic fields in terms of magnetic textures. Example one (Fig. \ref{fig:MnPtSn}) pertains to an antiskyrmion lattice stabilized in a nonstoichiometric Mn$_{1.4}$PtSn single crystal \cite{VirChemMat,Nayak2017,Jena2020, Jena2022, Peng2020}. Here, the thin electron beam transparent lamella cut from a single crystal is strained, which leads to ubiquitous bending contours from Bragg scattering in the image contrast (long stripes in Fig. \ref{fig:MnPtSn})(a) typical for single crystalline samples in the TEM. Thus, this example is suited to demonstrate the implications of intensity variations in an experimental TIE setting. The phase has been reconstructed along the same lines as the simulated examples above, with two deviations. Prior to the reconstruction, we applied an affine transformation of the defocused images in order to account for small magnification changes, rotation and distortion introduced by the defocus. Secondly, we employed a small Tikhonov regularization (see Appendix \ref{app: FDTIE}, regularization parameter $\lambda=125\,\mathrm{m}^{-1}$) to dampen some long-range phase artifacts (larger than 50\,nm) introduced by imperfectly fitted BCs (i.e., neglecting higher order Fourier coefficients) and the mildly ill-conditioned TIE. The comparison between finite-difference TIE (Fig. \ref{fig:MnPtSn}(c)) and the Poisson solution (Fig. \ref{fig:MnPtSn}(d)) shows that the former is much less affected by the bending contours (in particular at low-intensity regions in the lower right corner). Moreover, the Poisson solution exhibits again a general overestimation of the reconstructed field magnitudes. We note that non-trivial BCs affect a small region close to the boundary only because of the short characteristic length scale of the skyrmion lattice largely suppressing any impact of the boundary after one row of skyrmions. For comparison Appendix \ref{app: MnPtSn} also contains a simulation of the projected magnetic field in Mn$_{1.4}$PtSn lattice showing general agreement with the reconstructions.

\begin{figure}
\centering
\includegraphics[width=1\columnwidth]{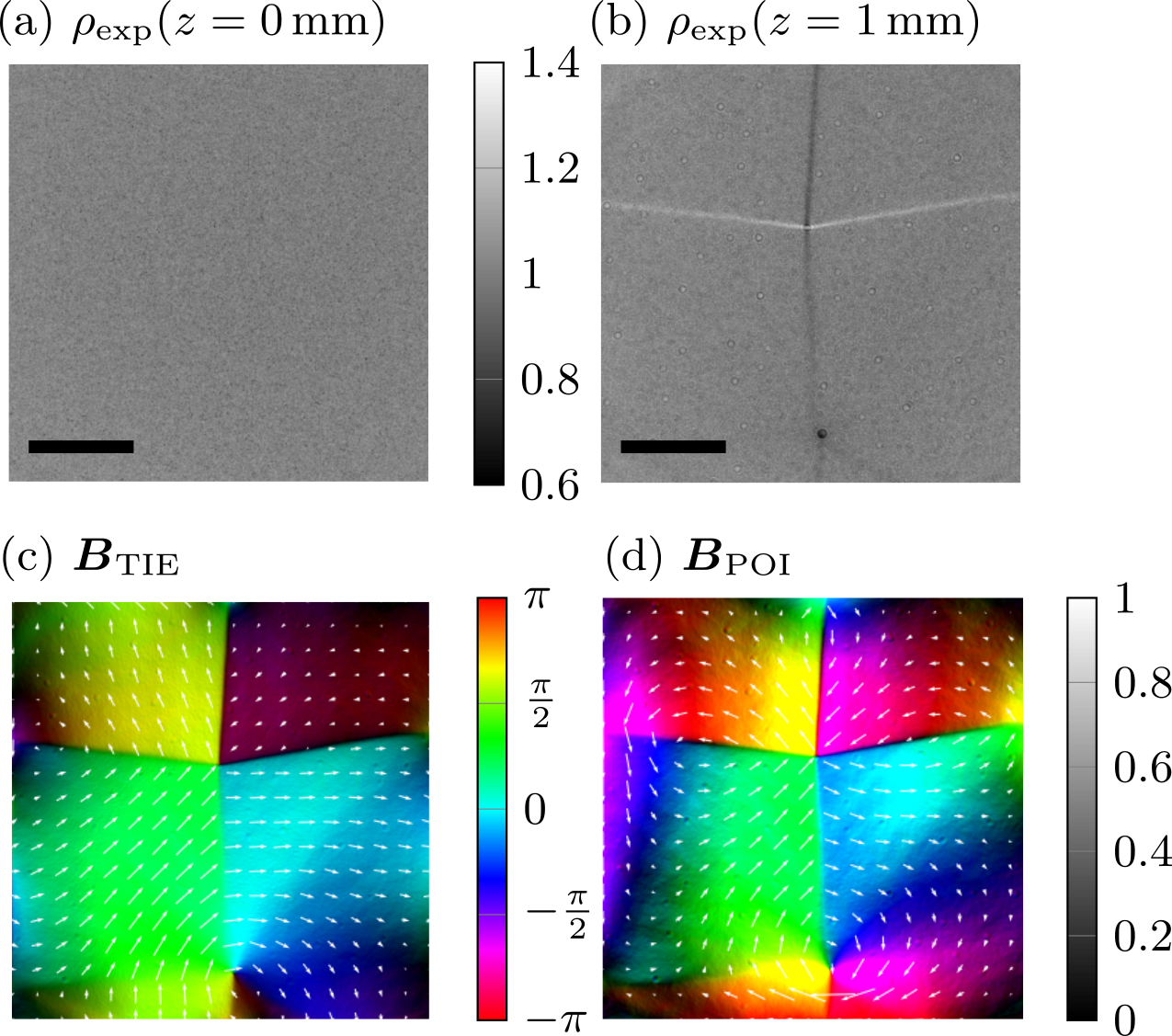}
\caption{\justifying TIE reconstruction of Landau pattern in a thin permalloy film: (a) in-focus TEM image; (b) TEM image at underfocus of 1mm;(c) magnetic field reconstructed with TIE; (d) magnetic field reconstructed with Poisson method. Scalebars correspond to 1.5\,\textmu m. The gray colormap encodes the normalized image intensity (a,b) and the normalized magnetic field magnitude (c,d), the hsv colormap denotes the angle of the magnetic field vectors (c,d).}
\label{fig:landau}
\end{figure}

The second example pertains to magnetic domains within a 30\,nm Permalloy (Py) film sputtered on a 2-5\,nm thin amorphous carbon film (Fig. \ref{fig:landau}). The latter is supported on a 50\,nm thick carbon square grid (Quantifoil S7/2) of 7\,µm period and 2\,µm bar width on a TEM specimen with hexagonal Cu grid. In this poly-crystalline film, practically no intensity variations are visible in focus (Fig. \ref{fig:landau}(a)). Due to the presence of a vortex and an antivortex the BCs are non-trivial, which allows demonstrating the BC variation of the improved algorithm in an experimental setting. Here, the comparison of the improved TIE (c) and the Poisson algorithm (d) shows that the former generally recovers the correct BCs, while almost preserving the constant in-plane magnetic field magnitude across the reconstruction domain, in stark contrast to the Poisson method. Stronger reconstruction artifacts can be noted in the finite-difference TIE reconstruction at positions, where the domain walls run into the boundary. These artifacts are again introduced by the limited number of Fourier coefficients (6 in this case) representing the boundary function, which cannot represent sharp cusps of the boundary function at the domain walls.

\section{Summary}

We have demonstrated that solving the TIE with the help of a finite difference scheme incorporating a variation of the boundary conditions largely mitigates reconstruction artifacts encountered with the previously used Poisson reconstruction schemes. The reconstruction algorithm reduces or even removes the need for additional regularization and mostly improves the reconstruction of long-range variations of the phase. That opens new avenues for TIE reconstructions, where intensity variations cannot be avoided, and long-range fields are important. A prominent example from magnetic imaging in the TEM would be vector field tomography, which requires large reconstruction domains containing the whole magnetic object and hence large intensity variations, in particular at the edge of the object, due to diffraction contrast.
Further improvements of the algorithm, e.g., concerning the optimization of the boundary conditions taking into account more Fourier coefficients and speed, hinge on further optimization of the optimization algorithm employed for determining the boundary conditions, faster sparse matrix solvers, and transition from a finite difference to a finite elements scheme (that allow adaptive meshing), amongst others.

\section{Acknowledgement}
This work was supported by the Deutsche Forschungsgemeinschaft DFG through the SFB 1143, project-id 247310070 and through the Würzburg-Dresden Cluster of Excellence on Complexity and Topology in Quantum Matter – ct.qmat (EXC 2147, project-id 390858490). M.W. acknowledges support from the International Max Planck Research School for Chemistry and Physics of Quantum Materials (IMPRS-CPQM).

\bibliographystyle{apsrev4-2}
\bibliography{bib}

\appendix

\section{Finite Difference TIE} \label{app: FDTIE}
We interlace the $x$ and $y$ sampling points so that the phase and intensity at these points may be represented by a single vector $\boldsymbol{\varphi}$ and $\boldsymbol{\rho}$. Subsequently, we approximate the directional derivatives along $x$ and $y$, which are the building blocks of the nabla and divergence operators, by the right- and left-sided differences represented by the following matrices
\begin{equation}
    \mathbf{D}^{(r)}_x=\frac{1}{\delta x}\begin{pmatrix}
    -1 & 1 &  &  \\
     & \ddots & \ddots &  \\
     &  & -1 & 1 \\
     &  &  & -1     
    \end{pmatrix}\,
\end{equation}
\begin{equation}
    \mathbf{D}^{(l)}_x=\frac{1}{\delta x}\begin{pmatrix}
    1 &  &  &  \\
    -1 & 1 &  &  \\
     & \ddots & \ddots &  \\
     &  & -1 & 1     
    \end{pmatrix}\,,
\end{equation}
\begin{equation}
    \mathbf{D}^{(r)}_y=\frac{1}{\delta y}\begin{pmatrix}
    -1 & \boldsymbol{0}_{N_x-1} & 1 &  & \\
     & \ddots & \boldsymbol{0}_{N_x-1} & \ddots & \\
     & & \ddots & \boldsymbol{0}_{N_x-1} & 1 \\
     & & & \ddots & \\
     & & & & -1     
    \end{pmatrix}\,,   
\end{equation}
\begin{equation}
    \mathbf{D}^{(l)}_y=\frac{1}{\delta y}\begin{pmatrix}
    1 & & & & \\
     & \ddots & & & \\
    -1 & \boldsymbol{0}_{N_x-1} & \ddots & & \\
     & \ddots & \boldsymbol{0}_{N_x-1} & \ddots & \\
     & & -1 & \boldsymbol{0}_{N_x-1} & 1     
    \end{pmatrix}\,.    
\end{equation}
Here, $\boldsymbol{0}_{N_x-1}$ denotes a row vector of zeros of length $N_x-1$. The density factor stacked in between gradient and divergence is written as 
\begin{equation}
    \mathbf{P}=\begin{pmatrix}
    \rho_1 & & \\
     & \ddots & & \\
     & & \rho_{N_x N_y}    
    \end{pmatrix}\,.  
\end{equation}
With that the whole finite difference TIE reads
\begin{equation}
    \frac{\boldsymbol{\rho}\left(z+\delta z\right)-\boldsymbol{\rho}\left(z-\delta z\right)}{2\delta z} = -\frac{1}{k_0}\left(\mathbf{L}_x+\mathbf{L}_y\right)\boldsymbol{\varphi}
\end{equation}
with the modified second-order difference matrices
\begin{equation}
    \mathbf{L}_{x,y}=\frac{1}{2}\left(\mathbf{D}^{(l)}_{x,y}\mathbf{P}\mathbf{D}^{(r)}_{x,y}+ \mathbf{D}^{(r)}_{x,y}\mathbf{P}\mathbf{D}^{(l)}_{x,y}\right)\,.
\end{equation}
We note that we have symmetrized the latter to treat all boundaries (i.e., upper and lower, left and right) equivalently.
The linear system may now be regularized in the Tikhonov way by augmenting the corresponding variational problem by a $\lambda^2\boldsymbol{\varphi}^2$ penalty term, i.e.
\begin{equation} 
 \boldsymbol{\varphi}=\mathrm{argmin}\left(\|\frac{\boldsymbol{\rho}\left(z+\delta z\right)-\boldsymbol{\rho}\left(z-\delta z\right)}{2\delta z} + \frac{1}{k_0}\mathbf{L}\boldsymbol{\varphi}\|^2+\lambda^2\|\boldsymbol{\varphi}\|^2\right)\,,   
\end{equation}
with $\mathbf{L}=\mathbf{L}_x+\mathbf{L}_y$. Here $\lambda$ is the regularization parameter, which determines a high-pass filter radius in reciprocal space, when applying this regularization to the constant intensity (Poisson-like) solution (see below). The corresponding normal equation, i.e., the regularized TIE reads
\begin{equation} 
 -\mathbf{L}^\mathrm{T}\frac{\boldsymbol{\rho}\left(z+\delta z\right)-\boldsymbol{\rho}\left(z-\delta z\right)}{2\delta zk_0} = \left(\frac{1}{k_0^2}\mathbf{L}^\mathrm{T}\mathbf{L}+\lambda^2\mathbf{I}\right)\boldsymbol{\varphi}\,.   
\end{equation}
Applying this regularization to the $ \rho=\mathrm{const.}$  case (corresponding to Poisson equation) yields
\begin{equation}
\boldsymbol{\varphi}=\frac{k_0}{2\delta z\rho}\mathcal{F}^{-1}\left(\frac{q^2}{q^4+k_0^2\lambda^2}\mathcal{F}\left(\rho\left(z+\delta z\right)-\rho\left(z-\delta z\right)\right)\right)\,,
\end{equation}
which reveals the high-pass character of the Tikhonov regularization in this case, dampening low spatial frequencies below $\sqrt{k_0\lambda}$ in the reconstruction. We employ the (regularized) Poisson solver (in a duplicated domain to emulate homogenoeus pseudo Dirichlet BCs\cite{Volkov(2002)}) as a reference solution in the main text.

Arbitrary inhomogeneous BCs are incorporated into the above scheme by augmenting the discretized solution domain $\mathcal{D}$ with a pixel line around the whole domain. This boundary (denoted by $\partial \mathcal{D}$) is mapped to the discretized boundary function $\boldsymbol{f}$ via a set of equations that is integrated into the above linear equation system scheme 
\begin{equation}
    \boldsymbol{\varphi}_{\partial\mathcal{D}}=\boldsymbol{f}\,.
\end{equation}

\section{Fresnel Propagation of Reconstructed Wave} \label{app: FresnelProp}

Finding the correct BCs involves the computation of the intensities of the TIE reconstructed wave function in the defocused planes pertaining to the experimental data. To facilitate the numerical propagation of the wave function, we multiply the Fourier representation of the Fresnel propagator $\hat{T}_\mathrm{F}\left(\delta z\right)$
\begin{equation}
    \mathbf{T}\left(\boldsymbol{q}\right)=\mathrm{exp}\left(i \frac{\delta z}{2k_0}\boldsymbol{q}^2)\right)
\end{equation}
to the TIE wave function in Fourier space, which corresponds to a convolution in position space. In order to suppress boundary artifacts from the discrete Fourier transformation, we duplicate the whole domain, ensuring zero spill over from opposing sides of the reconstruction domain in that process. Alternatively, padding of the reconstructed wave with constants corresponding to the boundary values of the wave in a sufficiently large surrounding also works.

\section{Vortex and Antivortex Landau Pattern} \label{app: LandauPattern}
Landau patterns in a square geometry have been generated from four homogeneous magnetic domains, which are invariant along $z$ and are separated by domain walls. The domain walls have been assumed to be sharp with respect to the spatial resolution of the simulation. We also neglected the out-of-plane vortex / antivortex core modulation of the field. Assuming Coulomb gauge (i.e., $\nabla \cdot \boldsymbol{A}=0$) the corresponding vector potential in one domain, say the upper one in the vortex Landau pattern (Fig. 1), is non-zero in the $z$-component only and reads 
\begin{equation}
    A_z\left(x,y\right)=C\mu_0 y\,,
\end{equation}
where $C$ is a constant determining the magnitude of the magnetic field ($C=6400\,\mathrm{A}/\mathrm{m}$ in our case, which corresponds to the saturation magnetization of permalloy). The whole vector potential of the vortex / antivortex Landau patterns containing four domains has been obtained by rotating, shifting, and stitching the above single domain vector potential. The magnetic phase of the electron wave function may finally be obtained from the Aharonov-Bohm phase in axial scattering approximation $\varphi = -\frac{e}{\hbar}\int_0^t A_z dz=-\frac{e}{\hbar}\bar{A}_zt$. Consequently, average in-plane magnetic flux densities may be obtained from the phase via \begin{equation}
    \begin{pmatrix}
    B_x \\
    B_y 
    \end{pmatrix}=-\frac{\hbar}{et}\begin{pmatrix}
    \partial_y \bar{A}_z\\
    -\partial_x \bar{A}_z \end{pmatrix}\,.
\end{equation}

\section{Poisson-Type Solutions of TIE} \label{app: IrrCur}
\begin{figure}
\centering
\includegraphics[width=1\columnwidth]{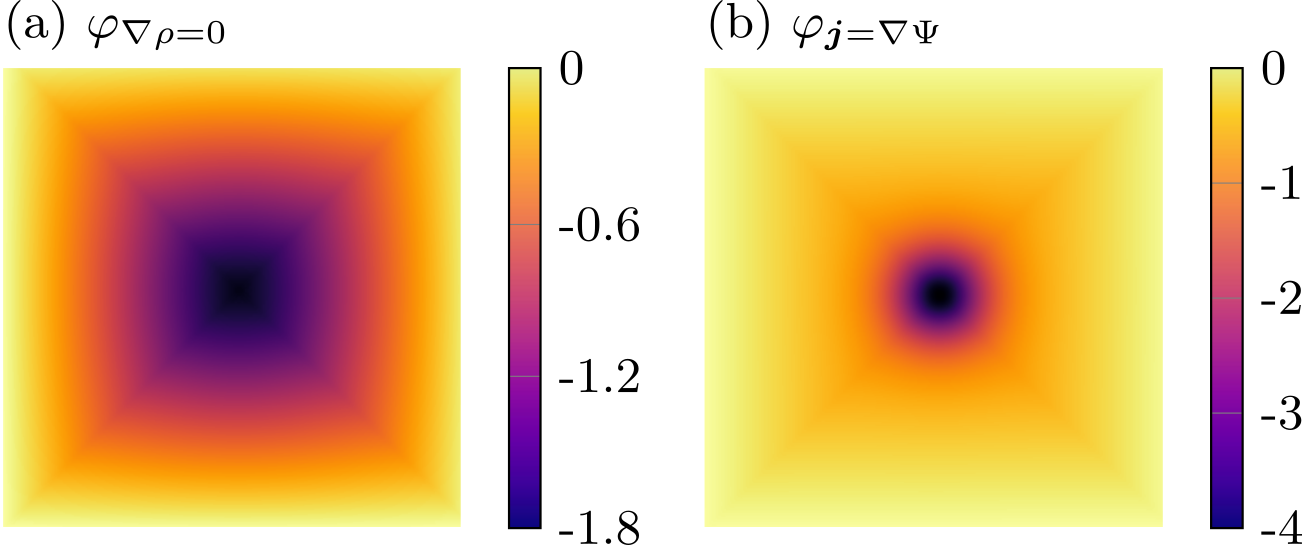}
\caption{\justifying Reconstructed phase of vortex Landau pattern associated to (a) $\nabla \rho=0$ approximation and (b) irrotational current.}
\label{fig:phase_irr}

\end{figure}

To approximate the exact solution of the TIE in the presence of variations of the image intensity $\rho$ by a solution of a Poisson equation several strategies can be employed. The most straight forward one consists of splitting the image intensity into its constant average and deviations from that average, i.e. $\rho\left(\boldsymbol{r},z\right)=\bar{\rho}\left(z\right)+\delta\rho\left(\boldsymbol{r},z\right)$. Using that expansion the TIE reads
\begin{equation}
\frac{\partial\rho}{\partial z} =-\frac{1}{k_0}\left(\bar{\rho}\Delta\varphi+\delta\rho\Delta\varphi+\nabla\delta\rho\cdot\nabla\varphi\right)  
\end{equation}
where we omitted the arguments to shorten the expression. In the main text we only kept the first term on the right hand side to compute the reference solution. Fig. \ref{fig:phase_irr}(a) displays the reconstructed phase pertaining to the vortex Landau pattern of the main text (Fig. \ref{fig:vortex}) also keeping the second term in above expansion, which is also frequently employed in literature. One observes a larger deviation from the true solution in this case, which, however, must not be considered a general rule. Depending on the structure of the solution and the intensity variation keeping the second term may also lead to improved solutions.

Finally, there is a third strategy, which also leads to a Poisson differential equation. The particle current defined by $\boldsymbol{j}=\rho \nabla \varphi$ may be generally decomposed into an irrotational (curl-free) and solenoidal (divergence-free) current (Helmholtz decomposition). Consequently, solving the Poisson equation
\begin{equation}
\frac{\partial\rho\left(\mathbf{r},z\right)}{\partial z} =-\frac{1}{k_0}\Delta\psi\left(\mathbf{r},z\right)
\end{equation}
generally corresponds to finding the scalar function $\psi$, whose gradient corresponds to the irrotational current. It has been proposed that this scalar function may be used to compute an associated phase by integrating $\nabla \varphi_\mathrm{irr} = \nabla \psi / \rho $.\cite{Teague(1983)} The corresponding Fourier space algorithm reads ($\tilde{\psi}=\mathcal{F}\left(\psi\right)$)
\begin{align}
\boldsymbol{\varphi}_\mathrm{irr}=&\frac{k_0}{2\delta z}\mathcal{F}^{-1}\Biggl(\frac{1}{q_x^2+q_y^2} \nonumber \\
&\left(q_x \mathcal{F}\left(\frac{\mathcal{F}^{-1}\left(q_x\tilde{\psi}\right)}{\rho}\right)+q_y \mathcal{F}\left(\frac{\mathcal{F}^{-1}\left(q_y\tilde{\psi}\right)}{\rho}\right)\right) \Biggr)\,,
\end{align}
where we again omitted the arguments to abbreviate the expression. Such a phase may significantly deviate from the original phase in case of intensity variations\cite{Zuo2014a}, which we demonstrate here at the vortex Landau pattern example discussed in the main text (Fig. \ref{fig:vortex}(b)). Fig. \ref{fig:phase_irr}(b) displays the results of the above reconstruction algorithm for the vortex example showing a strong deviation from the true phase (and hence the TIE solution) in the central region, where intensity variations have been large. 

As alluded to in the main text, the true application case of this solution is therefore the reconstruction of the irrotational part of the particle current, which is meaningful even in the presence of (partial) incoherence in contrast to the phase of a pure wave function.

\section{Magnetic Field Model of Mn$_{1.4}$PtSn}\label{app: MnPtSn}
\begin{figure}[h]
\includegraphics[width=1\columnwidth]{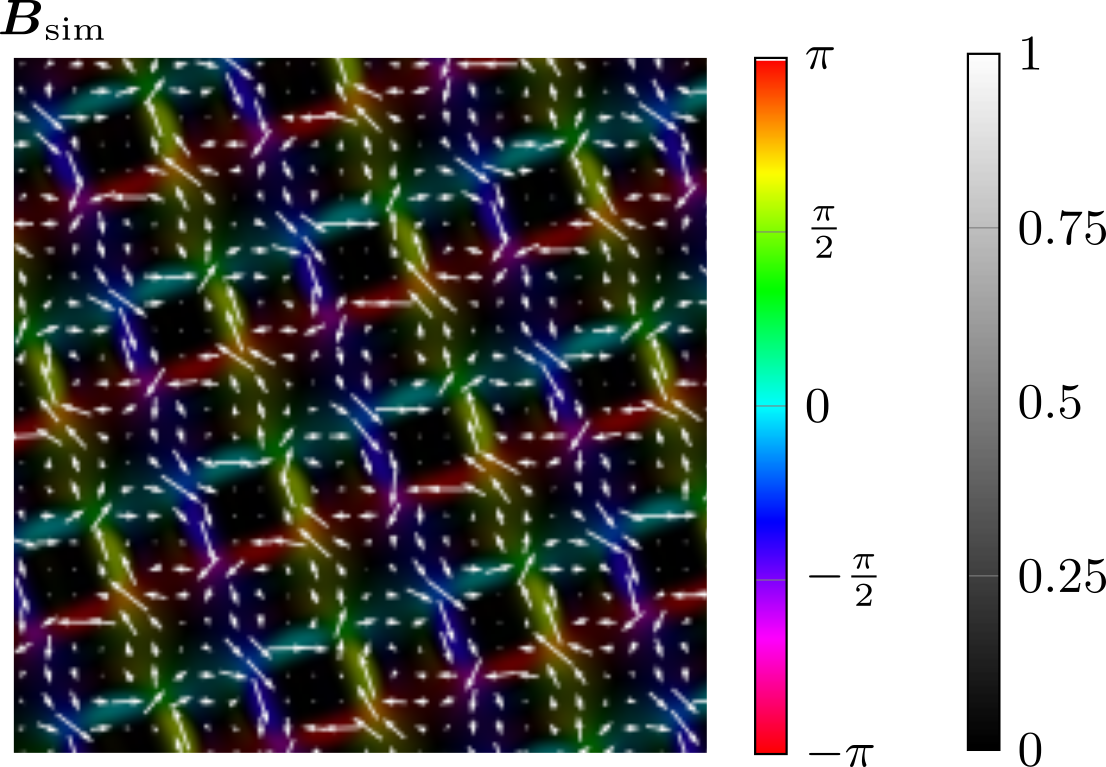}
\caption{\justifying Simulated projected in-plane component of the magnetic induction of antiskyrmion lattice in Mn$_{1.4}$PtSn.}
\label{fig:MnPtSnSim}
\end{figure}

As mentioned in the main text, we do not discuss the details of the reconstructed magnetic fields in this work. In order to provide minimal context, however, we supply a model projected magnetic induction for a thin (thickness = 100\,nm) slab of Mn$_{1.4}$PtSn, hosting a lattice of model antiskyrmions (Fig. \ref{fig:MnPtSnSim}). The simulation performed with Ubermag \cite{beg2022} using the micromagnetic parameters introduced by Peng et al. \cite{Peng2020} reveal similar field patterns (e.g., horizontal and vertical stripes) as reconstructed with the help of TIE. We, however, note also a couple of differences, which may be due (i) to the regularization error and remaining reconstruction errors, (ii) non-magnetic components in the phase (e.g., due to bending contours, thickness gradients) and (iii) deviations between the simulated and true magnetic structure.

\end{document}